\documentclass[pre,floatfix,aps,twocolumn]{revtex4-1}

\usepackage{subfigure}
\usepackage[utf8]{inputenc}
\usepackage{graphicx}
\usepackage{amsmath}
\usepackage{amsfonts}
\usepackage{bm}
\usepackage{color}
\usepackage{hyperref}

\hypersetup{colorlinks=true, citecolor=blue,
  urlcolor=black,
  pdfcreator={pdflatex},
}

\begin{document}
\title{Disordered hyperuniformity in two-component non-additive hard disk plasmas}

\author{Enrique Lomba$^{1,2}$, Jean-Jacques Weis$^3$, and Salvatore Torquato$^{2,4}$}
\affiliation{$^1$Instituto de Qu\'{i}mica F\'{i}sica Rocasolano,
CSIC, Calle Serrano 119, E-28006 Madrid,
Spain\\
$^2$Department of Chemistry, Princeton University, Princeton, New Jersey 08544, USA\\
$^3$Université de Paris-Sud, Laboratoire de Physique Théorique, UMR8627, Bâtiment 210,
91405 Orsay Cedex, France\\
$^4$Princeton Institute for the Science and Technology of Materials, Princeton University, Princeton, New Jersey
08544, USA
}

\begin{abstract}
We study the behavior of a two-component plasma made up of non-additive hard
disks with a logarithmic Coulomb interaction. Due to the Coulomb
repulsion, long-wavelength total density fluctuations are suppressed and the
system is globally hyperuniform. Short-range volume effects lead to
phase separation or to hetero-coordination for positive or negative
non-additivities, respectively. These effects compete with the hidden long-range order
imposed by hyperuniformity. As a result, the critical
behavior of the mixture is modified, with long-wavelength concentration
fluctuations partially damped when the system is charged. It is also
shown that the decrease 
of configurational entropy due to hyperuniformity originates from
contributions beyond the two-particle level. Finally, despite global
hyperuniformity, we show that in our system, the  spatial configuration associated with each
component separately is not hyperuniform, i.e., the system is not ``multihyperuniform."
\end{abstract}

\maketitle

\section{Introduction}
Disordered hyperuniform systems have gained considerable attention
over the last decade, since  their relevance as distinguishable states  of
matter was first stressed by Torquato and
Stillinger \cite{To03a}. Hyperuniform
many-body systems are those characterized by an anomalous
suppression of density fluctuations at long wavelengths
relative to those in typical disordered systems such as ideal gases,
liquids and structural glasses.  More precisely,
By definition,  a hyperuniform many-particle system in
$d$-dimensional Euclidean space $\mathbb{R}^d$ at number density $\rho$ is one
in which the structure factor $S({\bf Q})\equiv 1 + \rho {\tilde h}({\bf Q})$ tends to zero
as the wavenumber $Q \equiv |{\bf Q}|$ tends to zero \cite{To03a}, i.e.,
\begin{equation}
\lim_{Q\to 0} S({\bf Q}) = 0, 
\label{sq0}
\end{equation}
where ${\tilde h}({\bf Q})$ is the Fourier transform of the total correlation
function $h({\bf r}) = g_2({\bf r})-1$ and $g_2({\bf r})$ 
is the  pair correlation function.  

All perfect crystals and perfect quasicrystals, and certain special disordered systems are hyperuniform \cite{To03a,To15}. The fact that the microscopic
structure of disordered hyperuniform systems lie somewhere 
between that of disordered fluids (with only short-range disorder) and crystals (with long-range 
translational and orientational order) has been found to have
relevant consequences in a variety of contexts and applications  across different fields.
This includes maximally random jammed hard-particle packings \cite{Do05d}, classical disordered ground states \cite{Uc04b,To15,Zh17a,Zh17b}, 
driven nonequilibrium granular and colloidal systems \cite{He15,We15,Tj15}, dynamical processes in ultracold atoms \cite{Le14}, 
photonic band-gap materials \cite{Fl09b,Man13b,Fr17}, dense disordered transparent dispersions \cite{Le16},  photoreceptor mosaics in avian retina \cite{Ji14},  
immune system receptors \cite{Ma15}, composites with desirable transport, dielectric and fracture
properties \cite{Zh16b,Ch18,Xu17,Wu17}, polymer-grafted nanoparticle systems \cite{Chr17}, and ``perfect" glasses \cite{Zh17b}.

Our fundamental understanding of disordered hyperuniform systems is still 
in its infancy. We know that one can achieve them via equilibrium and
nonequilibrium routes, and they come in quantum-mechanical and classical
varieties. Classical disordered hyperuniform systems of identical particles in equilibrium necessarily
possess long-range interparticle interactions, whether they occur
at ground-state ($T=0$) conditions \cite{Uc04b,To15,Zh17a,Zh17b} or positive temperatures \cite{Ja81,Lo17a}.
However, much less is known about the hyperuniformity
of classical multicomponent systems and yet the infinite parameter
space (particle size distribution and composition) afforded by them
should provide greater tunablilty to achieve hyperuniform states.
Of course, when more than one component is present in the system, the situation
becomes obviously more involved, but also physically more
interesting, including technological relevance as designer composites \cite{Zh16b,Ch18,Xu17,Wu17}. 

It is noteworthy that when there are two or more components, the system can
be globally hyperuniform (long-wavelength total density fluctuations are
suppressed) or {\it multihyperuniform} (long-wavelength density fluctuations are
suppressed for each and every component). In practice,
multihyperuniformity in this case amounts to
suppressing simultaneously long-wavelength total density and concentration
fluctuations. These possibilities are accounted for by the Bhatia-Thornton structure
factors \cite{Bhatia1970}. In the first instance, one must consider the
total structure factor $S_{NN}(Q)$, defined as
\begin{equation}
  S_{NN}(Q) = \sum_{\alpha\beta} S_{\alpha\beta}(Q),
\end{equation}
where the partial structure factors, $S_{\alpha\beta}(Q)$ are given by
\begin{equation}
  S_{\alpha\beta}(Q) =
  x_\alpha\delta_{\alpha\beta}+\rho x_{\alpha}x_{\beta}\tilde{h}_{\alpha\beta}(Q),
  \label{spart}
\end{equation}
being $x_{\alpha}$ the mole fraction of component $\alpha$,
$\delta_{\alpha\beta}$ Kronecker's delta, $\rho$ the
total number density, and $\tilde{h}_{\alpha\beta}(Q)$ the Fourier
transform of the total partial correlation function
($h_{\alpha\beta}(r)=g_{\alpha\beta}(r)-1$, where $g_{\alpha\beta}$ is
the partial pair distribution function). The low-$Q$ behavior of the
total structure factor is connected to the isothermal compressibility of the systems, and it is known
to diverge when the critical point of a liquid-vapor transition is
approached (i.e., density fluctuations occur on the macroscopic length scale). 
A hyperuniform system is the antithesis of such a critical point with a total structure
factor that vanishes in this low-$Q$ limit according to (\ref{sq0}), but can be regarded to be at an ``inverted" critical
point in which the direct correlation function, defined
through the Ornstein-Zernike equation, becomes long-ranged \cite{To03a}.
Additionally, concentration fluctuations are described by the
concentration-concentration structure factor, defined by\cite{Bhatia1970,HansenBook2nd}
\begin{equation}
  S_{cc}(Q) = x_2^2S_{11}(Q)+x_1^2S_{22}(Q)-2x_1x_2S_{12}(Q)
  \label{scc}
\end{equation}
This quantity exhibits a low-$Q$ divergence when the binary system
approaches the consolute point, i.e. the demixing critical
point. Conversely, if the system is globally hyperuniform,
multihyperuniformity implies the suppression of low-$Q$ concentration
fluctuations, i.e.,
\begin{equation}
  \lim_{Q\rightarrow 0} S_{cc}(Q) = 0.
  \label{sqcclim}
\end{equation}

In this paper,  we theoretically and computationally
investigate  the behavior of two-component plasmas made up of non-additive hard
disks with a logarithmic Coulomb interaction. We will show that due to the Coulomb
repulsion, long-range total density fluctuations are suppressed and the
systems are globally hyperuniform at positive temperatures.
It is demonstrated that short-range volume effects lead to
phase separation or to hetero-coordination for positive or negative
non-additivities, respectively. Interestingly, we show that the decrease 
of configurational entropy due to hyperuniformity originates from 
contributions beyond the two-particle level. Finally, despite global
hyperuniformity, we show that in our system the  structure of each
component separately is not hyperuniform, i.e., the system is not ``multihyperuniform."

More specifically,  we will study one of the simplest disordered binary systems which
can exhibit hyperuniformity in two dimensions, namely, the
symmetric non-additive hard-disk (NAHD) plasma. This system is characterized by
a short range NAHD interaction, to which a long ranged repulsive
two-dimensional Coulomb interaction is superimposed. The
two-dimensional Coulomb
potential in plasma systems is known to lead to
hyperuniformity \cite{PRB_1978_17_2827,PhysRep_1980_59_1,Caillol1982},
with $\lim_{Q\rightarrow} S(Q) \propto Q^2$. On the other hand, the short range
part of the potential for positive non-additivity (i.e. when $\sigma_{\alpha\beta} >(\sigma_{\alpha\alpha}+\sigma_{\beta\beta})/2$,
being $\sigma_{\alpha\beta}$ the distance of minimum approach between
particles $\alpha$ and $\beta$), 
can induce a demixing transition\cite{Saija2002,Almarza2015}. In contrast, for negative
non-additivities the system will be fully miscible and presents a tendency to
hetero-coordination, i.e. local coordinations with neighboring unlike particles
tend to be favored. Our study makes extensive use of  Monte
Carlo simulations (MC) and integral-equation approaches, namely the
Hypernetted Chain (HNC) equation and the closely related Reference
Hypernetted Chain equation (RHNC). With these tools, 
we investigate the structural effects of the
interplay between long- and short-ranged interactions, with special
emphasis on the influence of hyperuniformity on the critical behavior
of the demixing transition for the NAHD plasma with positive
non-additivity. This is a particularly interesting situation, since
prior to demixing the system exhibits a structure reminiscent of
a disordered two-phase  heterogeneous material, which in this case will
be shown to be hyperuniform. 

The rest of the paper is organized as follows. In the next Section we
present our model system and provide a brief description of the
theoretical and simulation methods employed, including a summary of
the expressions that describe 
the system thermodynamics in the HNC approximation. Then, in Section
\ref{lowq}, we derive analytical expressions that describe the
low-$Q$ behavior of our system. It will be shown that
whereas the condition for global hyperuniformity is fulfilled, the
systems are not multihyperuniform. Finally, in Section \ref{res} we
present out most relevant results, with particular emphasis on the
phase behavior of the NAHD plasma and how global hyperuniformity affects the
concentration fluctuations that lead to demixing. The connection
between hyperuniformity, ``hidden order'', and configurational entropy
is also explored in this final Section.

\section{Model and methods}
Our model consists of a symmetric mixture of non-additive hard disks
with a two-dimensional Coulomb repulsion added, by which the
interaction potential is given by
\begin{equation}
  \beta u_{ij}(r) = \left\{
  \begin{array}{cc}
    \infty & \mbox{if}\; r < (1+\Delta(1-\delta_{ij})))\sigma \\
    - Z_iZ_j\Gamma \log r/\sigma
    \label{uijlog} & \mbox{if}\; r \ge (1+\Delta(1-\delta_{ij})))\sigma
  \end{array}
  \right.
\end{equation}
 where $\beta=1/k_BT$ as usual, $\Gamma =\beta e^2$, being $e$ the
 electron charge (esu units), $Z_i$, the particle charge in $e$ units
 (and here for simplicity we will just consider $Z_i=1$), $\Delta$ in the
 non-additivity 
 parameter, and  $\sigma$ the hard disk diameter between like
 species. Our system will be a mixture of total surface density,
 $\rho$. Theoretical calculations will be presented for the equimolar
 mixture $\rho_1=\rho_2=\rho/2$, i.e. the mole fractions will be  
 simply $x_1=x_2=1/2$. Throughout the paper density
 will be reduced as $\rho\sigma^2$. 

 \subsection{The integral equation approach}
 The Ornstein-Zernike equation for a mixture is given by \cite{HansenBook2nd}
 \begin{equation}
  \label{eq:oz}
  h_{jk} (r_{12}) = \sum_l \rho_l \int d{\bf r}_3 c_{jl} (r_{13})
  h_{lk} (r_{32}),
\end{equation}
where $c_{ij}$ is the  direct correlation function, with $\rho_l$
being the number density of species $l$. In
Fourier space Eq. (\ref{eq:oz}) can be cast into matrix form to yield
\begin{equation}
  \label{eq:ozft}
  \tilde{\bf{\Gamma}}(Q) = [{\bf I}-\tilde{\bf C}(Q)]^{-1}\tilde{\bf C}(Q)\tilde{\bf C}(Q),
\end{equation}
where ${\bf I}$ is the identity matrix, the tilde denotes a 2D Fourier transformation, and
\begin{eqnarray}
  [\tilde{\bf \Gamma}(Q) ]_{ij} &=&
  \sqrt{\rho_i\rho_j} \,\tilde{\gamma}_{ij}(Q), \\
  \protect[{\tilde{\bf C}}(Q)]_{ij} & = & \sqrt{\rho_i\rho_j} \,\tilde{c}_{ij}(Q),
\end{eqnarray}
with $\gamma_{ij} \equiv h_{ij}-c_{ij}$. The closure for Equation (\ref{eq:oz}) reads
\begin{equation}
  \label{eq:hnc}
  c_{ij}(r) = \exp\left[-\beta\phi_{ij} (r)+\gamma_{ij}(r)+B_{ij}(r)\right]-1-\gamma_{ij}(r).
\end{equation}
where $B_{ij}(r)$ is the bridge  function. Here we will make use of two
  approximations, namely $B_{ij}(r)=0$, i.e. the 
  HNC approximation, and $B_{ij}(r) = B_{ij}^{HS-PY}(r)$, where this
  latter function is the bridge function computed in the Percus-Yevick
  (PY) approximation for the plain NAHD system without electrostatic
  interactions. We will denote this approximation by RHNC-PY. These
  equations can now be solved by a simple iterative
  procedure, provided  the long range character of the correlations is
  appropriately treated. Following  Ref.\cite{JCP_2007_127_074501} we
  define a well-behaved long range component of the interaction
  \begin{eqnarray}
  \beta\phi_{ij}^{\rm LR}(r) & = & -Z_iZ_j \Gamma \left[\ln \left(\frac{r}{\sigma}\right) + \frac{1}{2}E_1\left(\frac{r^2}{\sigma^2}\right)\right], \\
  \beta\tilde{\phi}_{ij}^{\rm LR}(Q) & = &
  Z_iZ_j\Gamma\frac{2\pi}{(k\sigma)^2}\exp\left( -\frac{1}{4}(k\sigma)^2 \right),
  \label{eq:lr}
\end{eqnarray}
where $E_1(x)$ is the exponential integral. With these, one can
construct a set of short-ranged correlations and 
interaction of the form
\begin{eqnarray}
    \beta\phi_{ij}^{\rm SR}(r) & = & \beta\phi_{ij}(r)-\beta\phi_{ij}^{\rm LR}(r), \\
    c_{ij}^{\rm SR}(r) & = & c_{ij}(r)- \beta\phi_{ij}^{\rm LR}(r), \\
    \gamma_{ij}^{\rm SR}(r) & = & \gamma_{ij}(r)+ \beta\phi_{ij}^{\rm LR}(r),
\end{eqnarray}
and similarly for their Fourier transforms. Now, Equation (\ref{eq:ozft})
and its closure (\ref{eq:hnc}) can be solved without further
complications. A more detailed description of the procedure can be
found in \cite{JCP_2007_127_074501} and references therein. Here the
equations have been solved over 2000 grid points covering a range in
$r$-space of 20$\sigma$.

\begin{table*}
\begin{center}
\caption{ Thermodynamics of the equimolar NAHD plasma for
  $\rho\sigma^2=0.8$, and $\Delta=-0.2$, as computed from MC
  simulations, and in the HNC and RHNC-PY approximations.\label{thermo2}}
\begin{tabular}{cccccccc}
  \hline\hline
  & \multicolumn{1}{c}{MC} & \multicolumn{1}{c}{RHNC-PY} & \multicolumn{5}{c}{HNC} \\
$\Gamma$ & $\beta U^{ex}/N$ &  $\beta U^{ex}/N$ &  $\beta U^{ex}/N$ & $\beta A^{ex}/N$ & $S^{ex}/Nk_B$ & $S^{ex}_2/Nk_B$& $\Delta S^{ex}/Nk_B$\\
\hline
0.0  &   0     &  0      &  0      & 1.878 & -1.878 &-1.4520 & -0.426\\
0.5  & -0.248 & -0.2507 & -0.2407 & 1.651 &-1.892  &-1.4279 & -0.464 \\
1.0  & -0.512 & -0.5168 & -0.4990 & 1.405 &-1.904  &-1.4121 & -0.492 \\
2.0  & -1.053 & -1.0652 & -1.0318 & 0.897 & -1.929 &-1.3889 & -0.540 \\
5.0  & -2.733 & -2.7716 & -2.6840 & -0.688& -1.996 &-1.3456 & -0.650 \\
\hline \hline
\end{tabular}
\end{center}
\end{table*}

\begin{figure}[h]
  \includegraphics[width=8.5cm,clip]{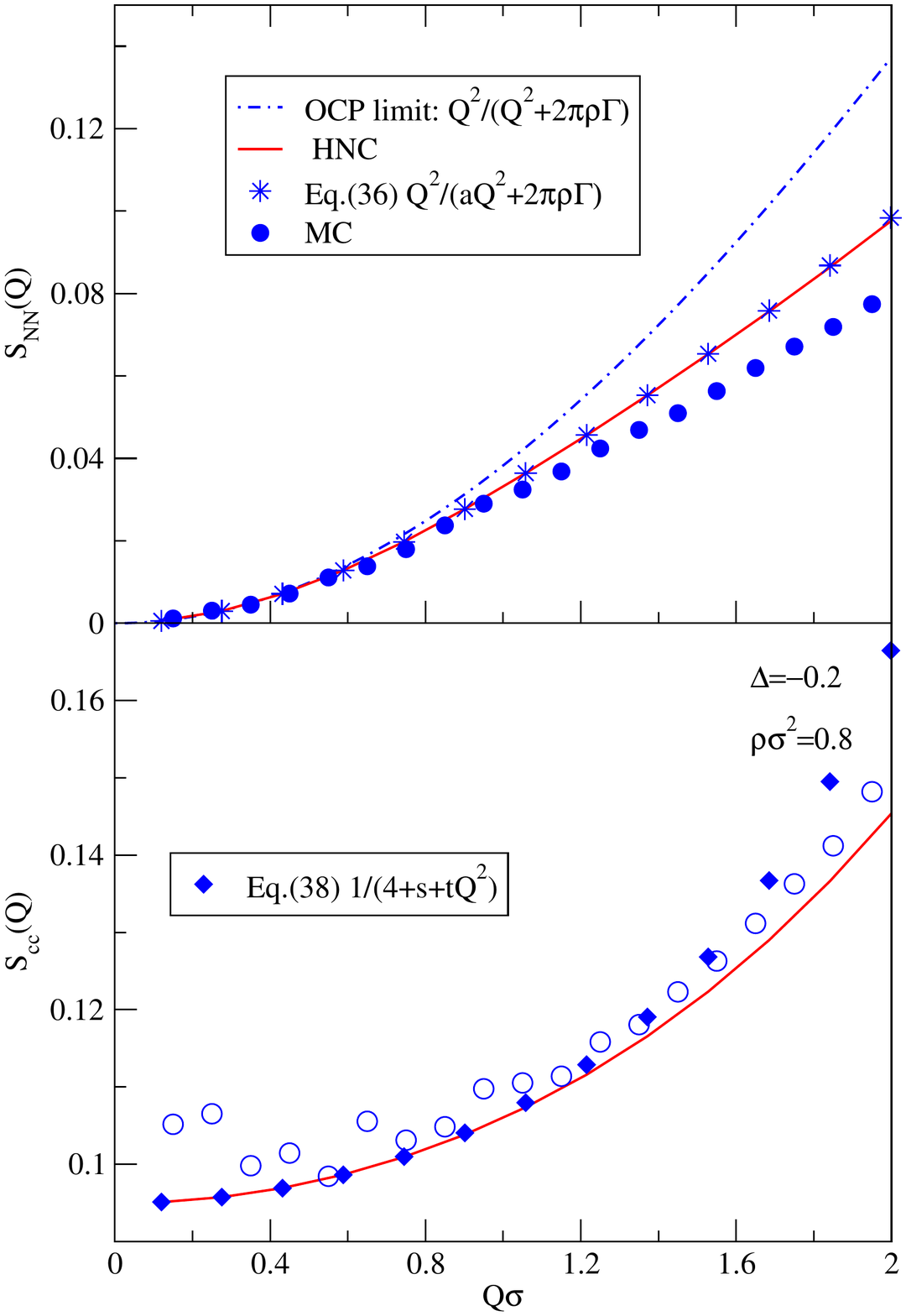}

\caption{Low-$Q$ behaviour of the $S_{NN}$ and $S_{cc}$ structure
  factors for an equimolar hard disk plasma mixture with negative
  non-additivity, as determined by simulation and in the HNC
  approximation. The low-$Q$ HNC expansion from Eq.~(\ref{snnlq}) and
  that the pure OCP limit ($a=1$) are also illustrated for $S_{NN}(Q$,
  together with the limiting formula (\ref{scclq}) for $S_{cc}(Q)$.\label{lowq08}}
\end{figure}

\subsection{Hypernetted Chain Equation thermodynamics}
\label{thermo}
In what follows, we summarize the key equations to compute the
thermodynamics of our system in the HNC approximation. Our choice is
based on the internal consistency of all thermodynamic properties in
the HNC (with the exception of the isothermal compressibility computed
from the
fluctuation theorem), which can be used to test the correctness of our results.
 The excess internal energy is simply
given by
\begin{equation}
  \beta U^{ext}/N = -\frac{1}{2\rho}\sum_{\alpha\beta}Z_\alpha Z_\beta \rho_\alpha\rho_\beta \Gamma \pi \int dr h_{\alpha\beta}(r) \log (r/\sigma).
\end{equation}
The accuracy of the internal
energy calculation, is high in the HNC, as can be appreciated in
Table \ref{thermo2}. Even though the RHNC-PY
provides slightly better results, here we will mostly make use of the HNC
thermodynamics, since quantities such as the chemical potential and
Helmholtz energy can be evaluated directly from the correlation
functions, and as mentioned above, it is endowed with a high degree of
thermodynamic consistency. Specifically, the  
Helmholtz's free energy can be evaluated as the sum of two contributions
\begin{equation}
  \beta A^{ex}/N =  \beta A^{ex}_1/N + \beta A^{ex}_2/N,
\end{equation}
obtained by a set of integrals in physical ${\bf r}$-space and
$Q$-space, namely
\begin{widetext}
\begin{eqnarray}
  \beta A^{ex}_1/N &=&
  \frac{\pi}{\rho}\sum_{\alpha,\beta}\rho_\alpha\rho_\beta\left(-\frac{\Gamma}{4}Z_\alpha
  Z_\beta +\int dr r \Big[
  c^{SR}_{\alpha,\beta}(r) +
  \frac{1}{2}[c^2_{\alpha,\beta}(r)-\gamma^2_{\alpha,\beta}(r)]\Big]\right)\nonumber\\
  \beta A^{ex}_2/N &=&-\frac{1}{2\rho}\int dQ Q \left(\log|{\bf I +
    \tilde{H}}(Q)/Q^{-1}| - {\mbox Tr}[{\bf
      \tilde{H}}(Q)/Q^{-1}]\right).
  \label{A1}
\end{eqnarray}
\end{widetext}

One can also calculate the free energy from
the chemical potential in the HNC, which is given by
\begin{widetext}
\begin{equation}
  \beta \mu_i^{ex} =
  -\sum_{\alpha}\rho_{\alpha}\tilde{c}_{i\alpha}^R(0) +
  \frac{\pi}{2}\sum_{\alpha}\rho_{\alpha}\int
  h_{i\alpha}(r)\left(h_{i\alpha}(r)-c_{i\alpha}(r) \right) r dr,
  \label{muex}
\end{equation}
\end{widetext}
where 
\begin{equation}
  \tilde{c}_{ij}^R(0)=\tilde{c}_{ij}^{SR}(0)+\frac{\pi}{2}\Gamma
  Z_{i}Z_{j}
\end{equation}
is the $Q\rightarrow 0$ limit of the regular part of the direct
correlation\cite{PhysRep_1980_59_1}. With this one gets,
\begin{equation}
  \beta A^{ex}/N = \sum_i\rho_i\beta\mu^{ex}- \beta P/\rho + 1.
\end{equation}
Now, the pressure can be computed from the virial equation, that in
the case if the NAHD two component plasma is simply
\begin{equation}
  \beta P/\rho = 1 + \frac{1}{4\rho}\pi \sum_{\alpha,\beta}
  \rho_\alpha\rho_\beta\sigma_\alpha\sigma_\beta g_{\alpha\beta}
  (\sigma_{\alpha\beta}^+)-\frac{1}{4}\Gamma.
  \label{vir}
  \end{equation}
The fact that the Coulomb contribution equals $-\Gamma/4$ can be used
as an internal consistency check of the results by numerically integrating the
correlation functions with the corresponding virial factors. 

From these expressions, the excess
configurational entropy is obtained as
\begin{equation}
  S^{ex}/Nk_B = \beta U^{ex}/N-\beta A^{ex}/N
\end{equation}
Additionally, one can determine the two-particle excess
configurational entropy contribution layer by layer using
\begin{eqnarray}
  S^{ex}_2(R)/Nk_B &=& -\pi \rho\sum_{\alpha,\beta}x_\alpha
  x_\beta  \label{s2ex}
\\
  & &\times\int_0^R \left(g_{\alpha\beta}(r)\log
  g_{\alpha\beta}(r)-g_{\alpha\beta}(r)+1\right)r dr\nonumber.
\end{eqnarray}
The total two-particle contribution corresponds to $S^{ex}_2(\infty)/Nk_B$.
This quantity accounts in most cases for more than 80 per cent of the total
configurational entropy\cite{PHYSA_1992_187_145}. We will see later 
that our case deviates from the standard behavior in regular
fluids. A relevant quantity in connection with the configurational
entropy, $S^{ex}$ and  its two particle contribution, $S^{ex}_2$, is
their difference $\Delta S^{ex}/Nk_B= S^{ex}/Nk_B-S^{ex}_2/Nk_B$. For
purely repulsive interactions, the presence of zeros in $\Delta
S^{ex}(\rho)/Nk_B$ has been correlated with the location of a
fluid-solid transition\cite{PHYSA_1992_187_145}. Finally, it is worth
mentioning that the two particle excess entropy is closely related 
with the $\tau$-order metrics parameter, which is a
measure of translational order \cite{To15},
\begin{eqnarray}
  \tau &=& \frac{\pi \rho}{D}\sum_{\alpha,\beta}x_\alpha
    x_\beta\int h_{\alpha\beta}(r)^2 r dr \nonumber \\ 
  &  =&
  \frac{1}{D}\sum_{\alpha,\beta}\int\left(S_{\alpha\beta}({\bf
    Q})-x_{\alpha}\delta_{\alpha\beta})(S_{\alpha\beta}({\bf
    -Q})-x_{\alpha}\delta_{\alpha\beta}\right)d{\bf Q}, \nonumber \\
  \label{tau}
\end{eqnarray}
where $D$ is a suitable characteristic length (e.g., correlation length). This expression is in fact
the multicomponent generalization of the $\tau$ order metric defined  in Ref. \cite{To15}.
Importantly, the $\tau$ order parameter 
defined by (\ref{tau}) is given in terms of quantities that are experimentally accessible. 
Note that a closely related order metric was defined in physical space in terms of $|h(r)|$ \cite{Tr00}, instead of
$h(r)^2$. Obviously, while this does not modify the qualitative behavior of the order
parameter, it does not have a corresponding representation in terms of the
structure factor.  Comparing
Eqs.(\ref{s2ex}) and (\ref{tau}), one sees that the latter can easily
be obtained from (\ref{s2ex}) by a small $h$-expansion of the integrand, with the sign
changed. Ordered systems (e.g. crystals) will give infinite $\tau$,
whereas for the ideal gas $\tau$ vanishes, as the
does the excess two-particle entropy.
  
%\subsection{Monte Carlo simulation}
\subsection{Details of the simulation procedure}
The Monte Carlo simulations were performed mostly in the canonical ensemble
using $N$ particles embedded in a uniform neutralizing background 
in a square box of side $L$ with periodic boundary conditions. The energy of the
periodic system was evaluated by the Ewald summation method with
conducting boundary conditions: \cite{Leeuw1982} 
\begin{eqnarray}
\label{I2}
\beta U^{\rm{ex}} = &&  \frac{\Gamma}{4}  \sum_{i=1}^N \sum_{j=1}^N Z_i Z_j \sum_{\bf n} {}^{'}
   E_1\left(\alpha*^2 \left|\frac{{\bf r}_{ij}}{L}+ {\bf n}\right|\,\right) \nonumber \\
 &&  + \frac{\Gamma}{4 \pi} \sum_{{\bf n} \neq 0}  \displaystyle 
  \frac{e^{- \pi^2 {\bf n}^2/\alpha*^2}}{{\bf n}^2}
   {\bigg|\sum_{i=1}^N Z_i
   \exp\left(2 \pi i{\bf n} \cdot \frac{{\bf r}_i}{L}\right)\bigg|}^2 \nonumber \\
%   \exp(2 \pi i{\bf n} \cdot {\bf r}_i/L)\bigg|}^2 \nonumber \\
 &&  - \frac{\Gamma}{4}(\gamma + \ln \alpha*^2)\sum_{i=1}^N Z_i^2 +
 \frac{\Gamma \pi}{4 \alpha*^2}  \sum_{i=1}^N Z_i^2
   \nonumber \\
 &&  + \frac{\Gamma}{2} \ln \left(\frac{L}{\sigma}\right) \sum_{i=1}^N Z_i^2. 
\end{eqnarray}
In Eq.\ (\ref{I2}), ${\bf r}_{ij}={\bf r}_j-{\bf r}_i$,  
 and $\gamma = 0.5772156\ldots$  is Euler's constant.
The prime in the sum over 
${\bf n} = (n_x,n_y)$, with $n_x$,$n_y$ integers, restricts it to $i \neq
j$ for ${\bf n}=0$. The dimensionless parameter $\alpha^* =\alpha L$ controls the
relative contributions to the Ewald sum of the direct and reciprocal
space terms. With the choice  $\alpha^*=6$, adopted in our 
calculations, only terms with ${\bf n}=0$ need to be retained in the
first sum of  
Eq.\ (\ref{I2}). The sum in reciprocal space extends
over all lattice vectors ${\bf k} = 2 \pi  {\bf n}/L$ with $|{\bf n}^2|
\le 64$.   
The fourth term in Eq.\ (\ref{I2}) represents the interaction of the background
with itself.
In the simulations the Coulomb potential was scaled by $L$ rather than $\sigma$
as in the theoretical approaches. 
The last term in Eq.\ (\ref{I2}) was added to meet the theoretical choice.

In the vicinity of the demixing transition which occurs for positive
non-additivity of the disks we performed semi-grand canonical MC simulations
along the lines detailed in Ref. \cite{Almarza2015} for the plain NAHD system.
In particular, we took advantage of the cluster algorithm described in
Ref. \cite{Almarza2015} for identity sampling. Two particles of the same species are
considered linked within the same cluster when their separation is less than
$\sigma (1+ \Delta)$. With this choice cluster identity swaps do not lead to
particle overlaps and for the present symmetrical case where the chemical
potential difference $ \Delta \mu = \mu_A - \mu_B = 0$, the procedure leads to
a rejection free algorithm of composition sampling for a fixed set of particle
positions\cite{Almarza2015}.

Most simulations used 1600, 2500 and 3600 particles and generally structural
properties or order parameters were obtained  
 by averaging over
$3 \times 10^6$ to $5 \times 10^6$ trial translational moves per particle
after equilibration of the system. An identity swap was performed after
5 translational moves. Further calculations were carried out with 6100
and 8400 particles  to allow for a more precise determination of the
consolute point of the mixture.

\begin{figure}[h]
  \includegraphics[width=8.5cm,clip]{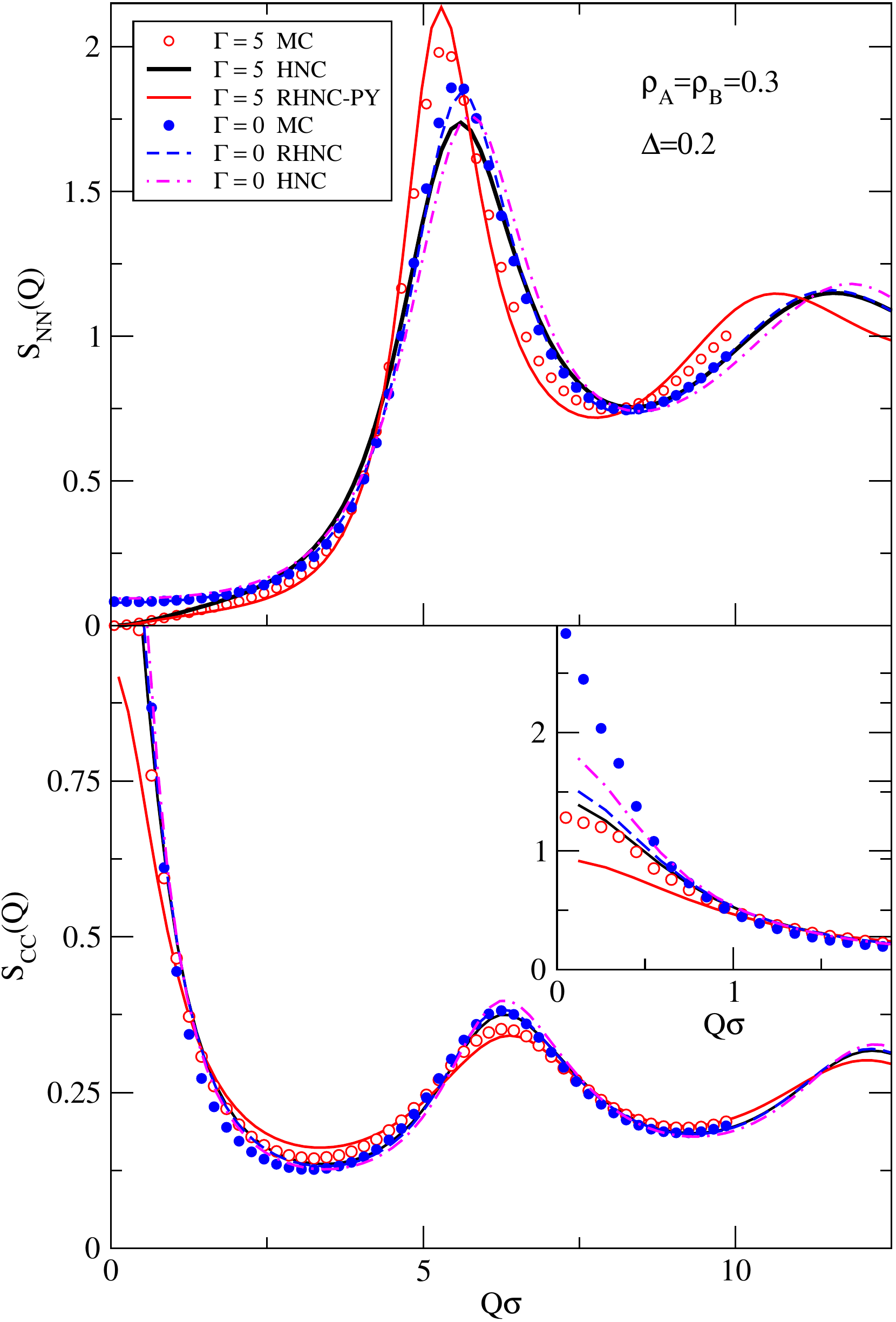}

\caption{Density-density, $S_{NN}$, and concentration-concentration,
  $S_{cc}$ structure factors for the equimolar NAHD plasma ($\Gamma=5$) and
  plain NAHD system $\Gamma=0$ for positive non-additivity. Curves
  denote various theoretical approaches (shown on the legend) and
  symbols MC data. The
  effect of global hyperuniformity is seen for the charged system, as
  $\lim_{Q\rightarrow 0}S_{NN}(Q)=0$.\label{sqfull}}
\end{figure}

\section{Low-$Q$ behavior}
\label{lowq}
Starting from the matrix form of the Ornstein-Zernike equation (\ref{eq:oz})
\begin{equation}
  {\bf I}+{\bf H} = \left[{\bf I} - {\bf C}\right]^{-1}
\end{equation}
one gets explicitly for the components of the partial structure
factors (\ref{spart})
\begin{eqnarray}
  1+\rho_i\tilde{h}_{ii} &=& \frac{1-\rho_j\tilde{c}_{jj}}{|{\bf I -
      C}|} \\
  \sqrt{\rho_i\rho_j}\tilde{h}_{ij} & = & \frac{\sqrt{\rho_i\rho_j}\tilde{c}_{ij}}{|{\bf I -
      C}|},
\end{eqnarray}
with $i\neq j$ and
\begin{equation}
  |{\bf I - C}| = 1 - \rho_1\tilde{c}_{11}
  -\rho_2\tilde{c}_{22}+\rho_1\rho_2(\tilde{c}_{11}\tilde{c}_{22}-\tilde{c}_{12}^2).
\end{equation}
The situation simplifies for the symmetric system, $c_{11}=c_{22}$, $\rho_1=\rho_2=\rho/2$. 
One can perform a small-$Q$ expansion of the direct correlation function
separating the Coulomb term,
\begin{equation}
  \tilde{c}_{ij}(Q) = \tilde{c}^R_{ij}(0)+c_{ij}^{(2)}Q^2-  2\pi\Gamma
  Z_iZ_j/Q^2.
\end{equation}
The expansion coefficients of
$\tilde{c}^R_{ij}(Q)$ 
are simply given by
\begin{equation}
  c_{ij}^{(2)}=\frac{1}{2}\left.\frac{\partial \tilde{c}_{ij}^R
    (Q)}{\partial Q}\right|_{Q=0}.
  \end{equation}
These $Q^2$ contributions are needed to reproduce the low-$Q$ behavior of the
partial structure factors beyond $Q=0$. 

\begin{figure}[b]
  \includegraphics[width=8.5cm,clip]{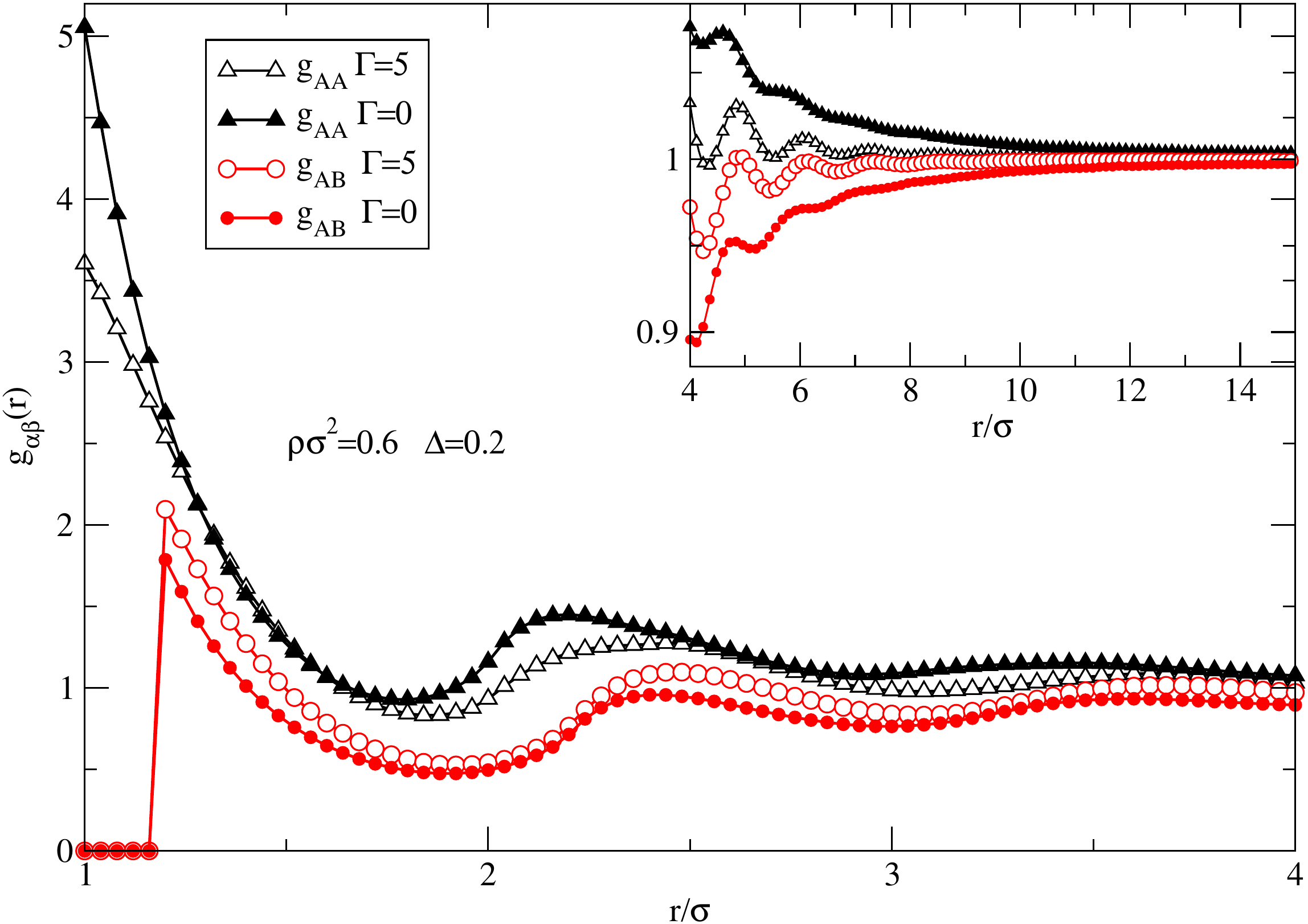}

\caption{Pair distribution functions  for the equimolar NAHD plasma ($\Gamma=5$) and
  plain NAHD system $\Gamma=0$ for positive non-additivity as obtained
  from MC simulation. The inset illustrates the long-range
  behavior. Long-range correlations are damped by the presence of
  charged, as an effect of global hyperuniformity .\label{grd06}}
\end{figure}

Using the expressions above, one can obtain the following limiting behavior
\begin{eqnarray}
1+\rho_1\tilde{h}_{11}(Q) &\approx &
  \frac{(1-\frac{\rho}{2}(\tilde{c}_{11}^R(0)+c_{11}^{(2)}Q^2))Q^2+\pi\rho\Gamma Z^2
    }{(aQ^2+2\pi\rho\Gamma Z^2)b}\nonumber\\
  \rho\sqrt{x_1x_2}\tilde{h}_{12}(Q) &\approx & \frac{\frac{\rho}{2}(\tilde{c}_{12}^R(0)+c_{11}^{(2)}Q^2)Q^2-\pi\rho\Gamma Z^2
    }{(aQ^2+2\pi\rho\Gamma Z^2)b},\label{h12l1}
\end{eqnarray}
when $Q\rightarrow 0$, with
\begin{eqnarray}
  a & = & 1 -\frac{\rho}{2}(\tilde{c}_{11}^R(0)+\tilde{c}_{12}^R(0)+(c_{11}^{(2)}+c_{12}^{(2)})Q^2)\nonumber
  \\
  b & = & 1 -\frac{\rho}{2}(\tilde{c}_{11}^R(0)-\tilde{c}_{12}^R(0)+(c_{11}^{(2)}-c_{12}^{(2)})Q^2).\nonumber
\end{eqnarray}

\begin{figure*}
   \center
   \subfigure[$\Gamma=0$]{\includegraphics[width=8cm,clip]{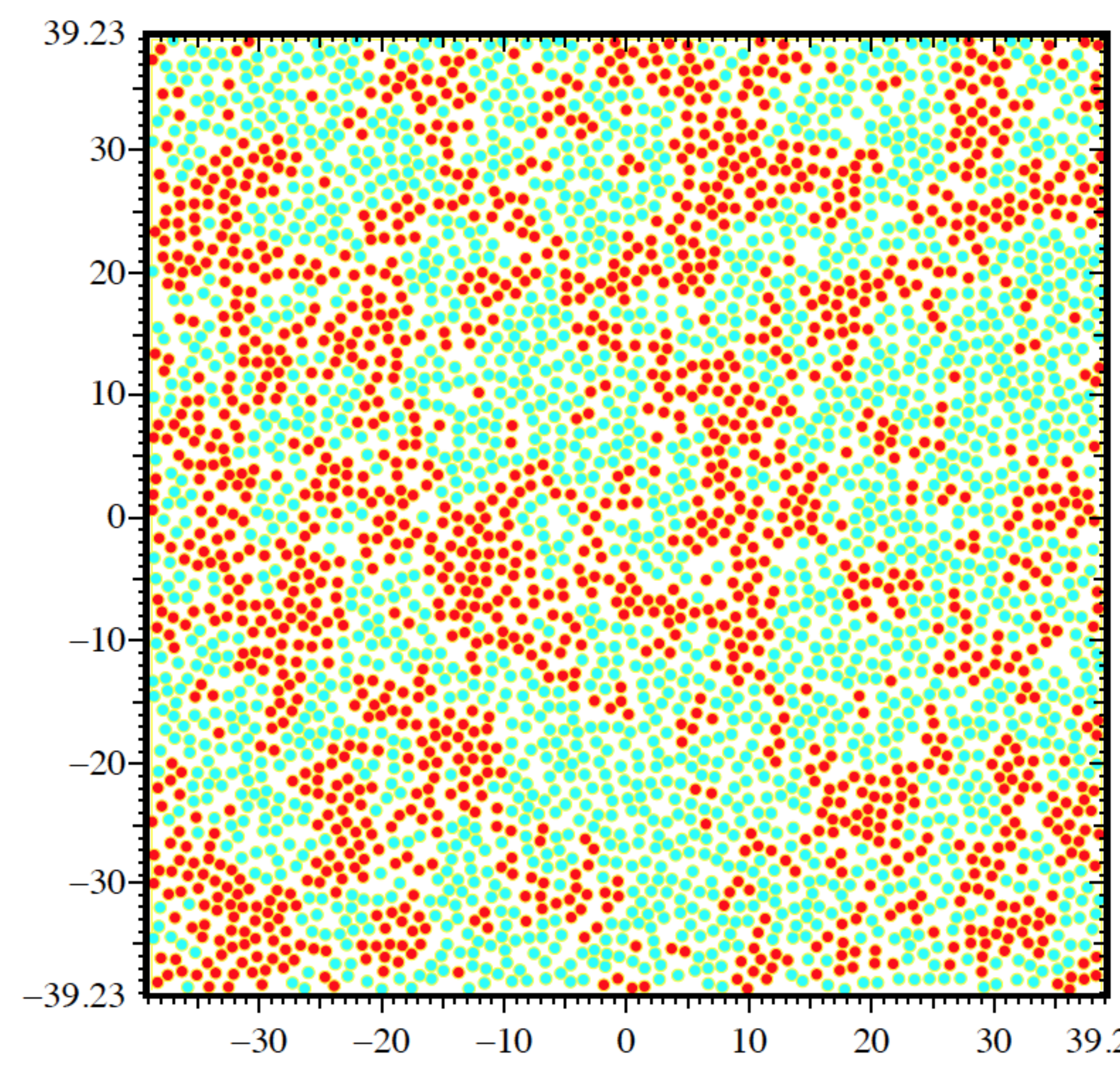}\label{snapg0}}
   \subfigure[$\Gamma=5$]{\includegraphics[width=8cm,clip]{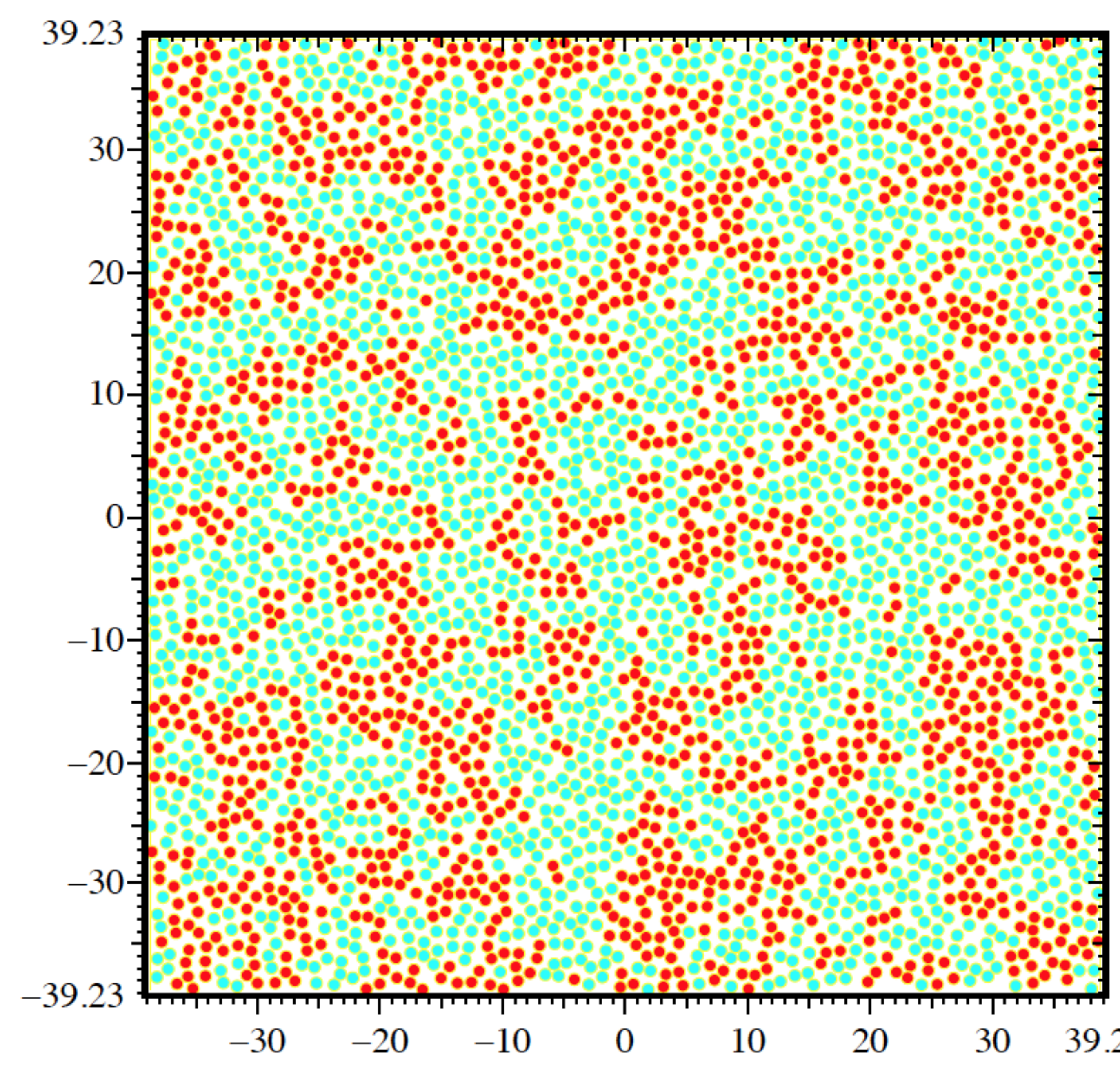}\label{snaplg5}}

   \caption{Snapshots of Monte Carlo configurations of the equimolar plain NAHD
     system ($\Gamma=0$) and the NAHD two
     component plasma  for $\Gamma=5$ and positive non additivity,
     $\Delta=0.2$, for $\rho\sigma^2=0.6$. The effect of global
     hyperuniformity cannot be appreciated on the translational order,
     but on the compositional order it can be seen that the presence
     of charges tends to reduce the size of the clusters.\label{snap}}
 \end{figure*}

For the total structure factor, in our fully symmetric mixture we have
\begin{eqnarray}
  S_{NN}(Q) &= &2S_{11}(Q)+2S_{12}(Q)
\end{eqnarray}
by which
\begin{equation}
  S_{NN}(Q) \approx \frac{Q^2}{aQ^2+2\pi\rho\Gamma Z^2}
  \label{snnlq}
\end{equation}
This differs slightly from the pure one component plasma (OCP)
behavior\cite{Caillol1982} for which $a=1$, since core contributions
vanish. In our case
non-negligible contributions from the hard core become more
evident as $Q$ grows.

For the concentration-concentration structure factor we have,
\begin{equation}
  S_{cc}(Q) =   \frac{1}{2}(S_{11}(Q)-S_{12}(Q)),
\end{equation}
which leads to
\begin{eqnarray}
  S_{cc}(Q)&\approx & \frac{1}{4b}=\frac{1}{4(1
    -\frac{\rho}{2}(\tilde{c}_{11}^R(0)-\tilde{c}_{12}^R(0)+(c_{11}^{(2)}-c_{12}^{(2)})Q^2))} \nonumber \\
&=&\frac{1}{4+s+tQ^2}
  \label{scclq}
\end{eqnarray}
where the constants $s$ and $t$ depend on density and on the short
range behavior of the direct correlation functions.

From the expressions (\ref{h12l1}), it is clear that
\begin{eqnarray}
  \lim_{Q\rightarrow 0}(1+\rho_1\tilde{h}_{11}(Q)) & = & \frac{1}{2(1 -\frac{\rho}{2}(\tilde{c}_{11}^R(0)-\tilde{c}_{12}^R(0)))}\nonumber\\
  \lim_{Q\rightarrow 0}(\rho\sqrt{x_1x_2}\tilde{h}_{11}(Q)) & = & -\frac{1}{2(1 -\frac{\rho}{2}(\tilde{c}_{11}^R(0)-\tilde{c}_{12}^R(0)))}\nonumber
\end{eqnarray}
which shows definitely that the system is not multihyperuniform,
although globally it has a hyperuniform behavior given by
Eq.(\ref{snnlq}), i.e. $S_{NN}(Q)\propto Q^2$ ($Q\rightarrow 0$). In
Figure \ref{lowq08} the validity of expressions  (\ref{snnlq}) and
(\ref{scclq}) is illustrated for a NAHD plasma with negative
non-additivity. For comparison the OCP limiting behavior is also
shown, and can be seen to deviate already at $Q\sigma \sim 0.6$. At
this point is it  important to stress that the global hyperuniformity
summarized in Eq.~(\ref{snnlq}) is the result of the symmetry
relation
\begin{equation}
  \lim_{Q\rightarrow 0}
  \left(\tilde{u}_{11}(Q)+\tilde{u}_{22}(Q)-2\tilde{u}_{12}(Q)\right)
  = 0
  \label{sym}
 \end{equation}
  being fulfilled by the long-range contributions to the
  interactions. Obviously, repulsive Coulomb systems comply with
  Eq.~(\ref{sym}) whenever $Z_1=Z_2$. 

\section{Results}
\label{res}
We have first focused our investigations in a case of positive non-additivity,
$\Delta=0.2$, whose phase behavior has already been studied in detail
for the uncharged system\cite{Almarza2015}. This NAHD mixture is known
to exhibit a demixing transition with Ising 2D criticality. Additionally, we have also
considered the situation with negative non-additivity, $\Delta =-0.2$,
which is characterized by the absence of a demixing transition and a
tendency to present local hetero-coordination.

\begin{table}[b]
\begin{center}
\caption{ Thermodynamics of the equimolar NAHD plasma for $\rho\sigma^2=0.6$ and $\Delta=0.2$ computed in the HNC approximation.\label{thermo1}}
\begin{tabular}{cccccc}
  \hline\hline
$\Gamma$ &  $\beta U^{ex}/N$ & $\beta A^{ex}/N$ & $S^{ex}/Nk_B$ & $S^{ex}_2/Nk_B$& $\Delta S^{ex}/Nk_B$\\
\hline
0.0  &  0      & 2.301 & -2.301 &-1.9947 & -0.306\\
0.5  & -0.2140 & 2.097 & -2.311 &-1.9484 & -0.362 \\
1.0  & -0.4423 & 1.879 & -2.321 &-1.9210 & -0.400 \\
2.0  & -0.9125 & 1.429 & -2.342 &-1.8813 & -0.461 \\
5.0  & -2.3677 & 0.030 & -2.398 &-1.8054 & -0.593 \\
\hline \hline
\end{tabular}
\end{center}
\end{table}

\subsection{Positive non-additivity and demixing transition}
In Ref.~\cite{Almarza2015}, it was found that the uncharged system
exhibits a phase separation when $\Delta=0.2$ with a consolute point
at  $\rho_c\sigma^2=0.69$ (and obviously $x_1=x_2=1/2$). We
have first studied the system at a somewhat lower total
density, $\rho\sigma^2=0.60$ and a relatively large Coulombic coupling,
$\Gamma=5$. Thermodynamic properties for this system in the HNC
approximation are collected in
Table \ref{thermo1}.

\begin{figure}[h]
  \includegraphics[width=8.5cm,clip]{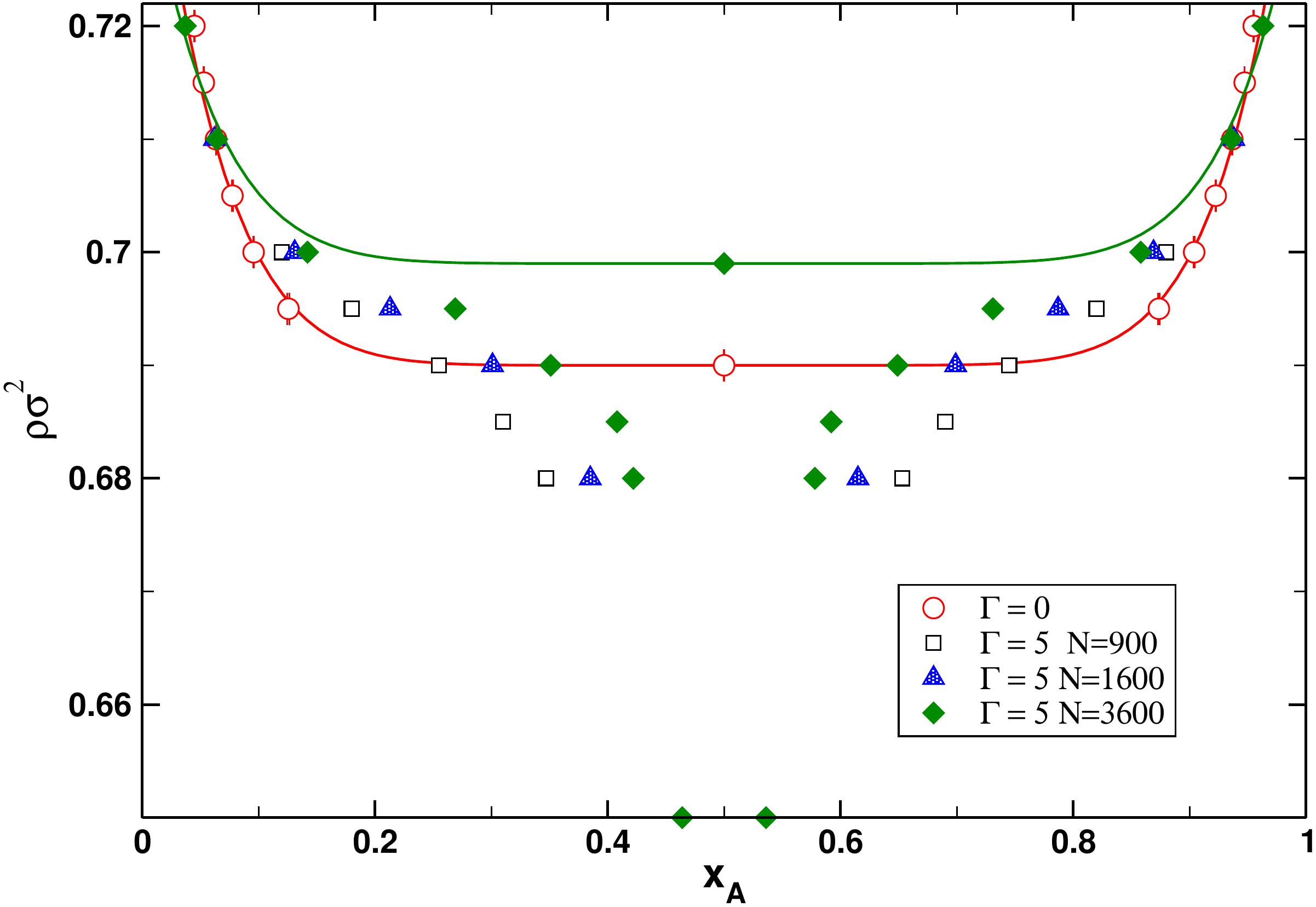}

\caption{Phase diagram of the NAHD system and the NAHD plasma with
  $\Gamma=5$. Symbols correspond to Semi Grand Ensemble simulations
  for different system sizes. The critical point estimates are
  obtained from the crossing of Binder's $U_{2n}$ cumulants (see
  Eq.~(\ref{u2n})) and the lines are a fit to Ising 2D critical behavior.\label{phase}}
\end{figure}

The effect of the charges is readily seen in the
density-density correlations represented by the total structure factor
(see Figure \ref{sqfull}), which now vanishes as $Q\rightarrow 0$,
making the system globally hyperuniform. On the other hand, in the
lower graph of Figure \ref{sqfull} for $S_{cc}(Q)$ we can observe that
there are large 
concentration fluctuations when $Q\rightarrow 0$, i.e. for large
separations. This is a clear indication of the vicinity of the
demixing transition. Interestingly, one observes that
$S_{cc}(0)(\Gamma=5) < S_{cc}(0)(\Gamma=0)$, i.e., charges (or global
hyperuniformity) counteract to a certain extent the tendency to
demix. This effect is further illustrated by the long-range behavior
of the pair distributions depicted in Figure \ref{grd06}. One readily
observes that the long-range oscillations of like and
unlike $g_{\alpha\beta}$ in the uncharged system (an indication of the
approaching divergence at the critical density) are quite damped due
the effect of the charges. Overall, one sees that the  values of the
like correlations at short and
intermediate ranges are lowered when charges 
are introduced, which reflects the repulsive nature of the Coulomb
interaction in the plasma. In contrast unlike correlations grow, 
since the Coulombic repulsion has a larger effect on
like particles whose distance of closest approach is $\sigma$, which
is lower than $(1+\Delta)\sigma$, for unlike particles. One can easily
see in the snapshots of
Figure \ref{snap} that this translates into a situation for which the
size of the clusters of
like particles decreases when charges are present. Here we have one of
these situations in which 
global hyperuniformity leads to some sort of long-range ``hidden''
order invisible to the eye. In contrast, the effect on the compositional
order is readily appreciated.

As to the net effect on the phase behavior, in Figure \ref{phase} we
present the phase diagram for the plain NAHD system, taken from
Ref.~\cite{Almarza2015} and that obtained in this work for $\Gamma=5$.
\begin{figure}[h]
  \includegraphics[width=8.5cm,clip]{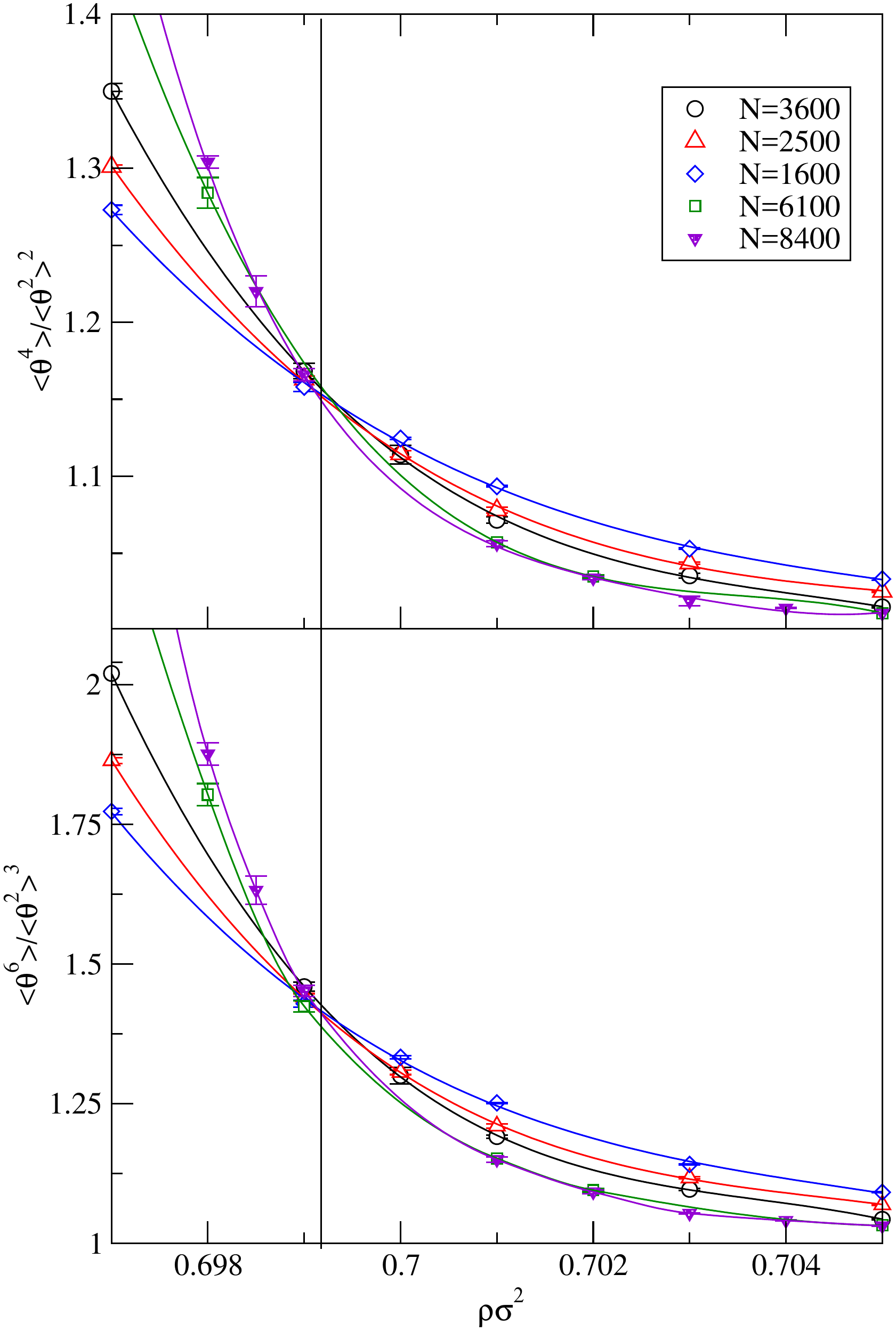}

\caption{Size dependence of Binder's $U_4$ and $U_6$ cumulants as a
  function of total density. The
  crossing for all curves is indicated by a vertical line and is seen
  to occur in the same density, $\rho_c\sigma^2\approx 0.699$.\label{Un}}
\end{figure}

The critical point estimates are calculated from the crossings of
Binder's cumulants\cite{landau-binder_book_2005}, $U_4$ and $U_6$,
\begin{equation}
  U_{2n} = \frac{\langle \theta^{2n}\rangle}{\langle\theta^2\rangle^n}
  \label{u2n}
  \end{equation}
where the $\langle\ldots\rangle$ denotes an ensemble average, and
$\theta=2x-1$. The size dependence of these quantities is illustrated
in Figure \ref{Un}, from which one can estimate the critical density
determined by the crossing of the curves. One obtains
$\rho_c\sigma^2\approx 0.699$, slightly larger than the value for the
plain NAHD system,  $\rho_c\sigma^2\approx 0.69$. This agrees
 with our previous findings that indicated that global
hyperuniformity, damping long-range correlations, tends to counteract
phase separation. Still, short range volume effects 
cannot be completely canceled out by the subtle changes induced by
global hyperuniformity and the system demixes at a higher density. As to the crossing of the cumulants,
for $U_4$, one gets $U_4^c\approx 1.15\pm 0.02$, which means that
the accepted Ising 2D universal value\cite{JSP_2000_98_551},
$U_4^c\approx 1.168$ lies within the uncertainty of our calculation. Here then, for our purposes, we have assumed 2D
Ising criticality\cite{JSP_2000_98_551}, and thus the fitted curves of Figure \ref{phase}
are obtained using a critical exponent $\beta=1/8$.

 \begin{figure*}
   \center
   \subfigure[$\Gamma=0$]{\includegraphics[width=8cm,clip]{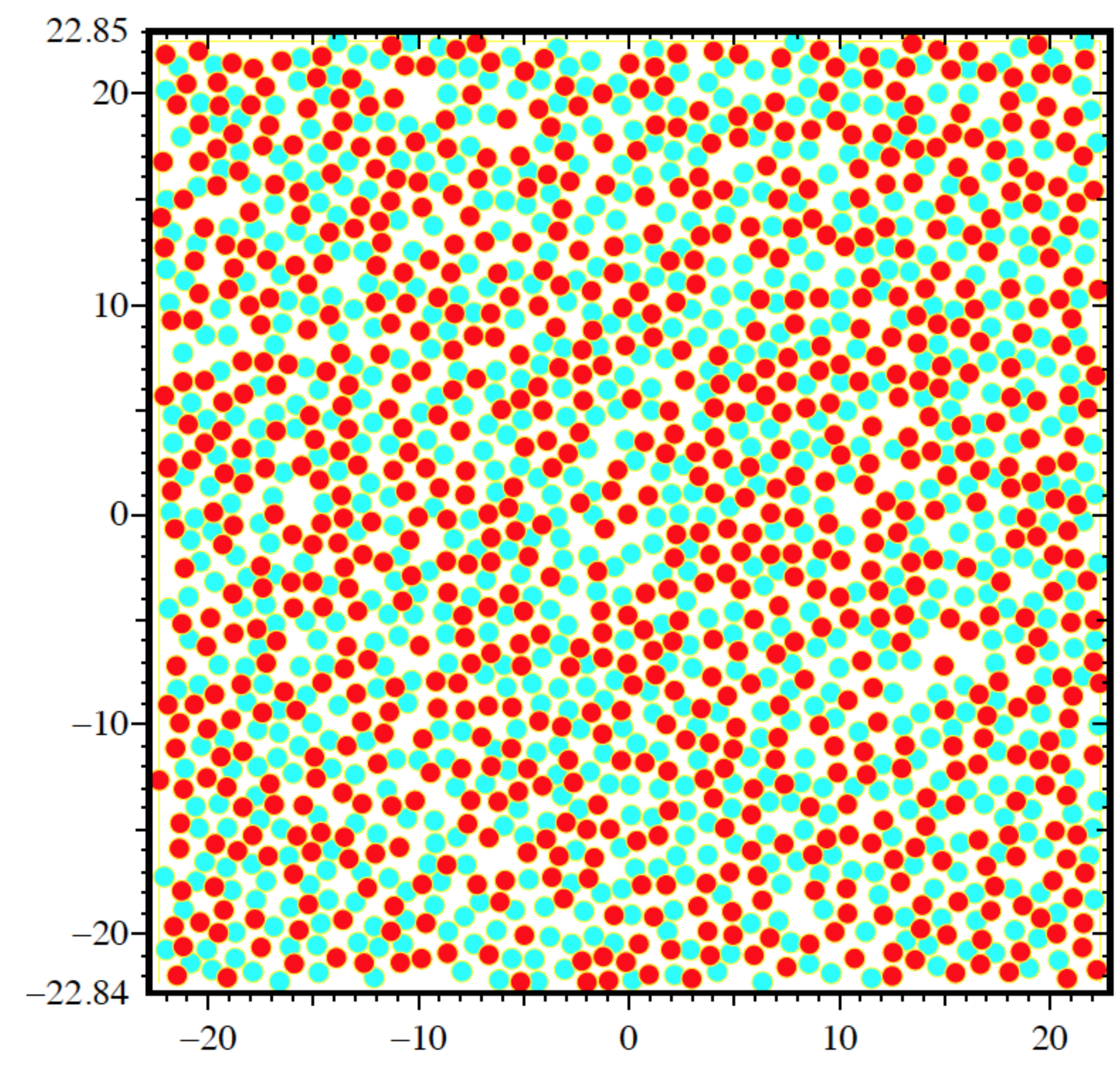}\label{snapug0}}
   \subfigure[$\Gamma=5$]{\includegraphics[width=8cm,clip]{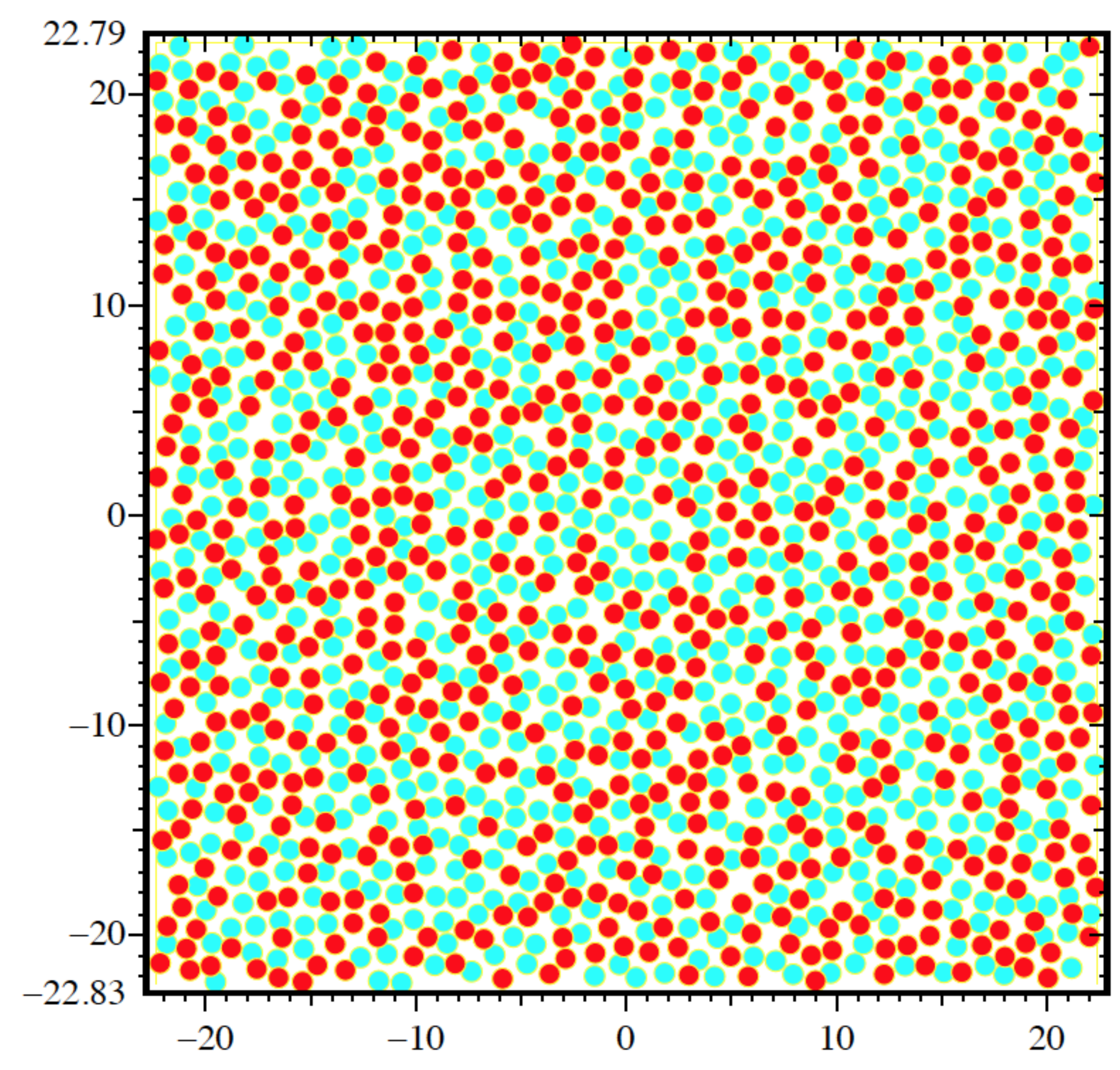}\label{snaplug5}}

   \caption{Snapshots of Monte Carlo configurations of the equimolar plain NAHD
     system ($\Gamma=0$) and the NAHD two
     component plasma  for $\Gamma=5$ and negative non additivity,
     $\Delta=-0.2$for $\rho\sigma^2=0.6$. The system tends to
     hetero-coordination, but the net Coulombic repulsion somewhat
     enhances clustering of like particles.\label{snap08}}
 \end{figure*}

 \subsection{Negative non-additivity: hetero-coordination}

\begin{figure}[h]
  \includegraphics[width=8.5cm,clip]{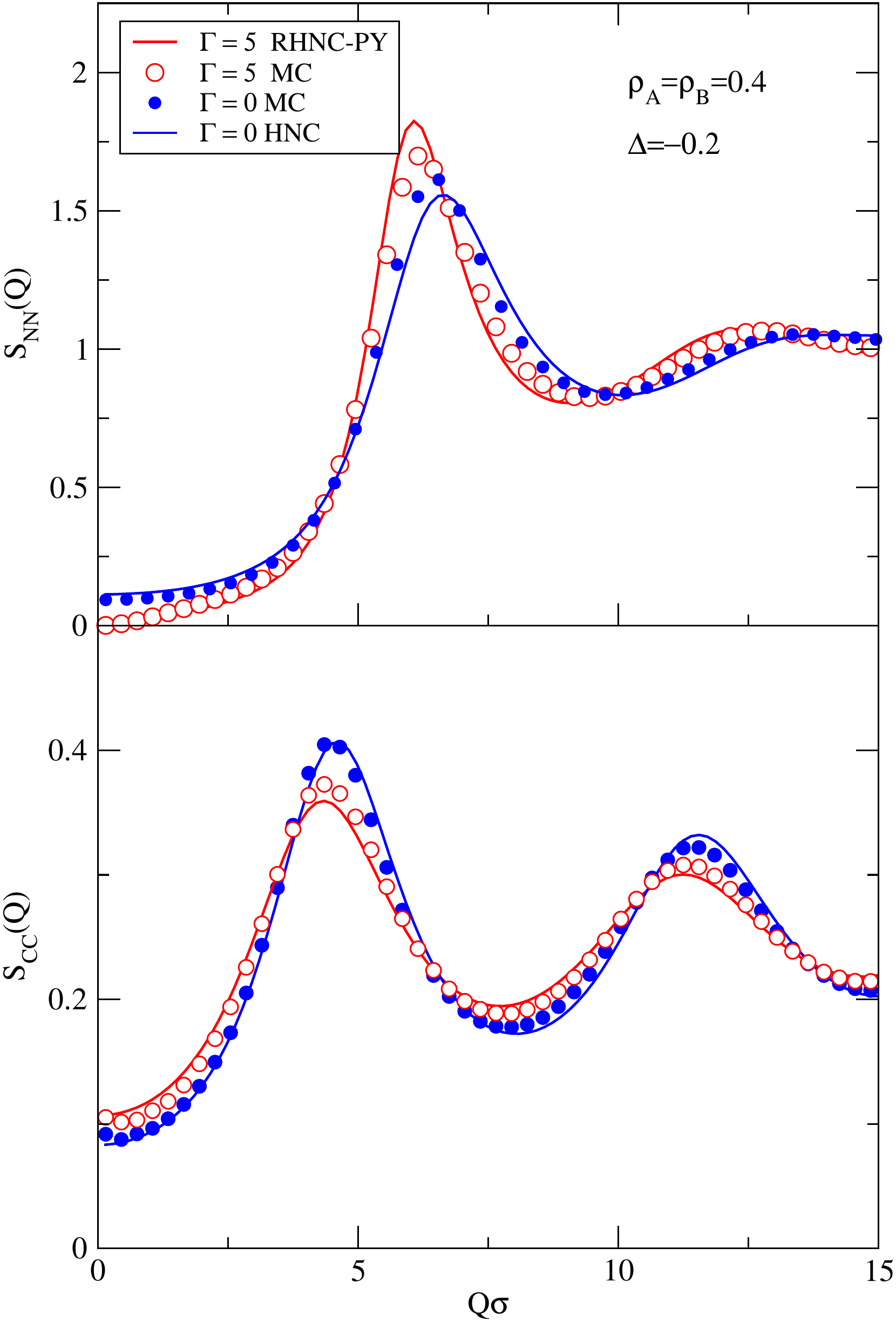}

\caption{Density-density, $S_{NN}$, and concentration-concentration,
  $S_{cc}$ structure factors for the equimolar NAHD plasma ($\Gamma=5$) and
  plain NAHD system $\Gamma=0$ for negative non-additivity. Curves
  denote various theoretical approaches (shown on the legend) and
  symbols MC data. Again, the effect of global hyperuniformity is seen
  for the charged system, as $\lim_{Q\rightarrow 0}S_{NN}(Q)=0$.\label{sqd08}.}
\end{figure}

 Thermodynamic properties for the NAHD system with $\rho\sigma^2=0.8$
 and $\Delta=-0.2$ are collected in Table \ref{thermo2} for various
 couplings. As to the structure, in the snapshots of 
 Figure \ref{snap08} one can qualitatively appreciate the effects
 of the charges (i.e. global hyperuniformity) on the microscopic
 structure of the fluid. On the left (uncharged NAHD) one can see that
 system tends to hetero-coordination, maximizing the contacts between
 unlike particles and thus minimizing volume. Switching on the
 Coulombic repulsion, even though it affects all particles in the same
 degree, has  more apparent  effects for pairs of unlike particles. In
 this case,
 their hard core repulsion allows for closer contact, and as a
 consequence, hetero-coordination is no longer so favorable. One
 can then appreciate a slight increase of like particle ``aggregates''.
\begin{figure}[b]
  \includegraphics[width=8.5cm,clip]{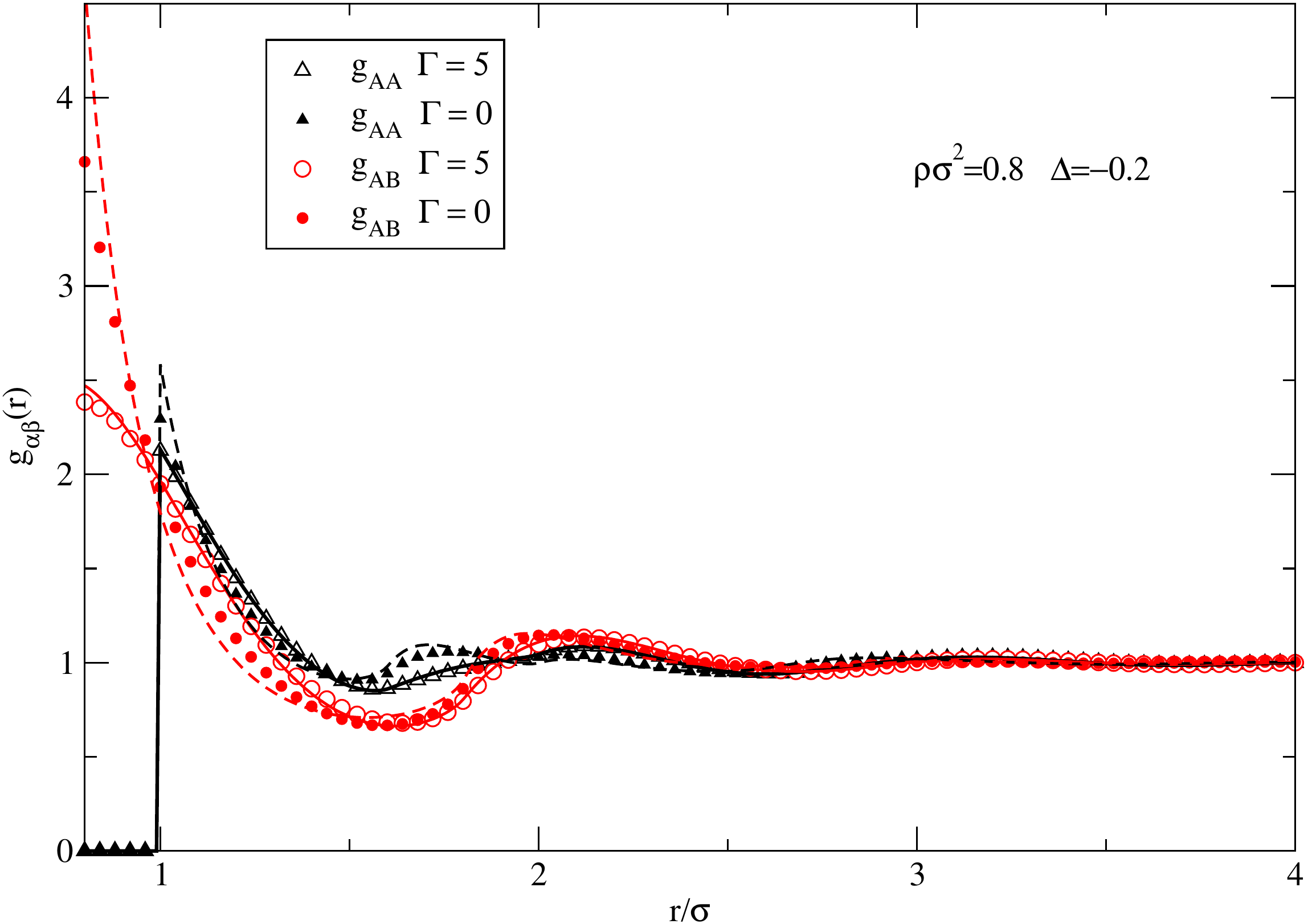}

\caption{Pair distribution functions  for the equimolar NAHD plasma, ($\Gamma=5$), and
  plain NAHD system, $\Gamma=0$, for negative non-additivity as obtained
  from MC simulation (symbols) and RHNC-PY approximation
  (curves).\label{gr08}}
\end{figure}
As a matter of fact, this is quantitatively illustrated by the
behavior of the partial pair distribution functions depicted in Figure
\ref{gr08}. There one can appreciate the considerable drop in the
contact value of the unlike pair distribution function (in contrast
with the situation for positive non-additivity seen in Figure
\ref{grd06}). Moreover, the like distribution function also decreases
somewhat, although to a much lesser extent. For this reason, the
snapshot of Figure \ref{snaplug5} seems to present a certain degree
of clustering of like particles. This features translates into a total
structure factor that decays to zero following (\ref{snnlq}) (i.e. a
globally hyperuniform system), and a concentration-concentration
structure factor with a low-$Q$ behavior given by Eq.~(\ref{scclq}),
as discussed in Section \ref{lowq} and illustrated in Figure
\ref{lowq08}. Additionally, in Figure \ref{sqd08}, we observe 
long-ranged oscillations in $S_{cc}(Q)$ (with a period of
$Q_l\approx 7\sigma^{-1}$) that actually reflect the
presence of hetero-coordination (i.e changes in local concentration
over a range $2\pi/Q_l\approx 0.9\sigma\approx
\sigma_{\alpha\beta}$). The presence of charges somewhat damps the
oscillations, i.e. it counteracts the hetero-coordination induced by
volume effects.

\begin{figure}[h]
  \includegraphics[width=8.5cm,clip]{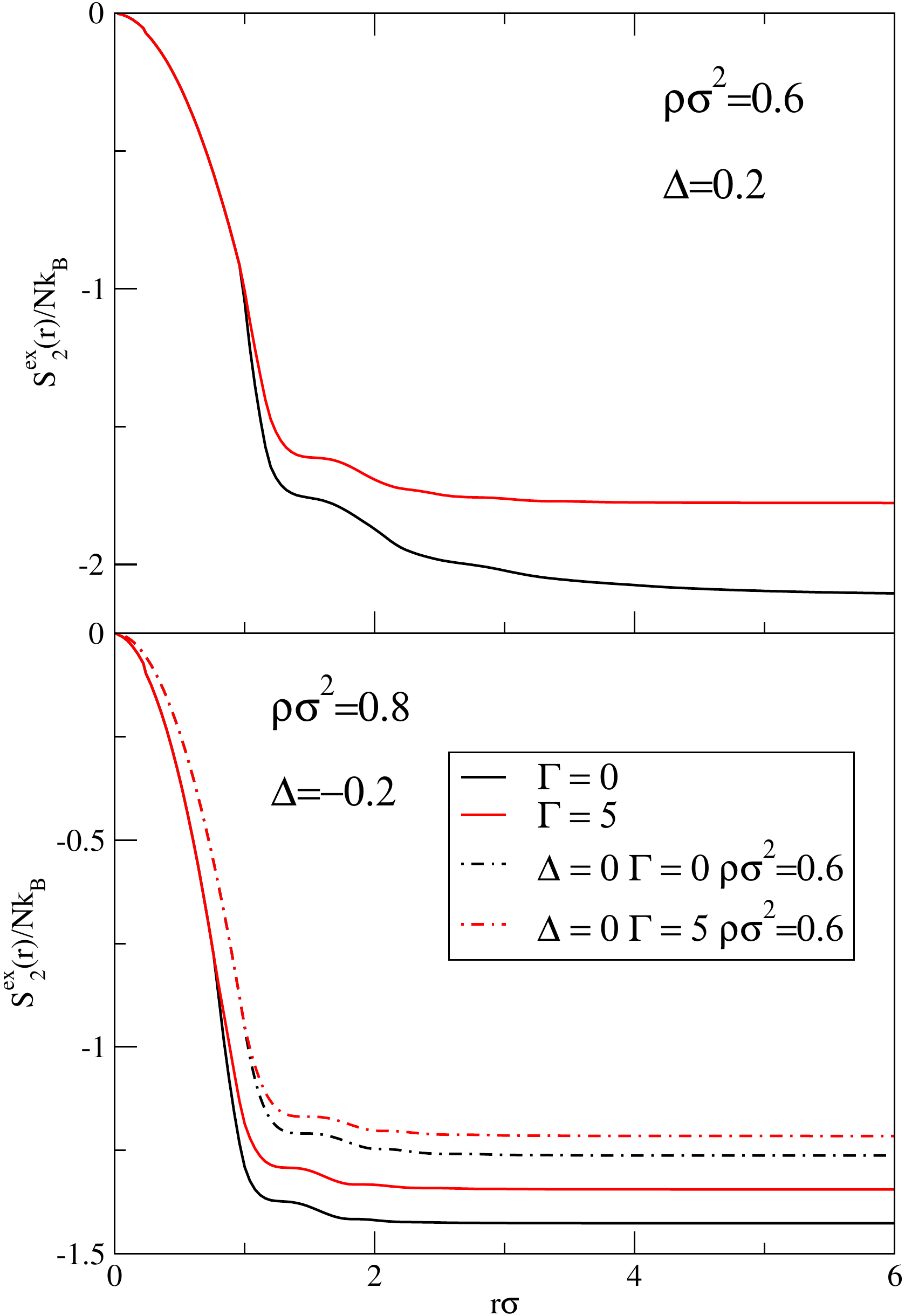}

\caption{Two-particle contribution to the excess configurational
  entropy determined in the HNC approximation for  equimolar NAHD fluids, with
  (black curves)
  and without (red curves). Dash-dotted curves correspond to an
  additive hard sphere system with $\rho\sigma^2=0.6$ (which gives a
  hard core contribution to the pressure similar to $\rho\sigma^2=0.8$
  with $\Delta=-0.2$).\label{s2_ex}}
\end{figure}

\subsection{Entropy and hyperuniformity}
As mentioned before, the two particle contribution to the
configurational entropy (\ref{s2ex}) usually accounts for 80\% of the
total configurational entropy. One can actually estimate the different
contributions of each particle layer from $S^{ex}_2(R)/Nk_B$, and thus
analyze the effect of charges on disorder (or more properly, on the
number of configuration/microstates compatible with our thermodynamic
state). In Figure \ref{s2_ex} we have plotted this quantity for both
the positive  and negative non
additivity computed in the HNC approximation. Additionally, in the
lower graph we have included the results for an additive system
($\Delta=0$), where no volume effects are at play. In this latter
instance, we chose a density that gives a hard core contribution to
the pressure similar to that of the $\Delta=-0.2$ case. One should
expect to a find an obvious effect on the two-particle entropy due to
the ``hidden order'' introduced by hyperuniformity. We observe that
except in the uncharged system close to demixing, the two-particle excess
entropy contributions originate in the first two coordination
shells. For $\Gamma=0$ and $\Delta=0.2$ this extends up to $4\sim 5$
layers. Interestingly, in
all instances one observes that the two-particle entropy decreases when charges are
introduced (i.e. when the system becomes hyperuniform). In the case of
negative non-additivity the effect is rather extreme. As mentioned,
the effect of hyperuniformity in this system counteracts phase
separation. Since the system looses entropy when demixing, it is
understandable that the charged system, further away from the
transition, might have a larger entropy. This
would explain why  the two-particle contribution is less
negative when charges are added. The situation is less obvious for
$\Delta=0$ and $\Delta=-0.2$. Actually, as discussed above, in Figure
\ref{gr08} one already sees that the net effect of the Coulomb
repulsion is a decrease in the contact values of the pair distribution
functions (much more visible in the unlike case). This could be
considered formally equivalent to the effect of a density decrease,
which obviously would imply an increase of entropy. This scenario
applies both to the additive and negative non additive hard disk
plasmas. At least, that is the situation as far as the two-particle
contribution is concerned. Note that as discussed in Section
\ref{thermo}, these arguments also apply to the evolution of the
translational order parameter, which is basically the negative of the
two-particle entropy.

Turning attention to the  the net configurational entropy, 
Tables \ref{thermo2} and \ref{thermo1} interestingly show that the increase
in entropy due to the two-particle configurational entropy
is overcompensated by many-particle contributions that are
approximated by the various terms that enter
Eqs. (\ref{A1})-(\ref{vir}). Now, one can vividly see a clear decrease of the
net configurational entropy. 
When comparing $S^{ex}/Nk_B$ and $S_2^{ex}/Nk_B$,   it is readily
seen  how the relative contribution of the two-particle
excess configurational entropy decreases as the system is charged, going from
80\% to 67\%. This is in marked contrast with the situation
for ``ordinary'' fluids where the two-particle contribution is known to
account for $80\sim 90$\% of the total configurational
entropy\cite{PHYSA_1992_187_145}. This reflects how
the ``hidden order'' introduced by hyperuniformity, being a long-range
effect, must influence entropy through many particle contributions: 
As we have seen, the two-particle configurational entropy is
determined basically by the first four coordination shells. 

In summary, we have shown that a simple system of NAHD with
superimposed repulsive two-dimensional Coulomb interactions can
exhibit a rich structural behavior due to the interplay between the
short range volume effects leading to phase separation, clustering or
hetero-coordination, and the long-range effects introduced by the
Coulomb forces inducing global hyperuniformity. Subtle effects are
particularly visible when hyperuniformity counteracts demixing. It is
worth mentioning that a completely similar picture would have been
obtained had the Coulomb interaction been fully three dimensional, with
the particles constrained to lie on a plane. The only relevant
difference would be a linear decay of $S_{NN}({\bf Q})$ as $Q\rightarrow
0$, instead of the quadratic dependence of Eq.(\ref{snnlq}).

Finally, we note  that elsewhere we have shown that by tuning interactions
in binary mixtures of non-additive hard-disk plasmas one can 
achieve disordered multihyperuniform many-body systems \cite{Lo17a}.   
We demonstrated that multihyperuniformity competes with phase separation and
stabilizes a clustered phase.

\acknowledgments
Prof. Giancarlo
Franzese is gratefully acknowledged for suggesting the analysis of the
entropy contributions.
E. L.  acknowledges the support from the Direcci\'on
General de Investigaci\'on Cient\'{\i}fica  y T\'ecnica under Grant
No. FIS2013-47350-C5-4-R, and  from the Program Salvador de
Madariaga, PRX16/00069 which supports his sabbatical stay at the
Chemistry Department of Princeton University. S. T. was supported by
the National Science Foundation under Award No. DMR-
1714722.

%\bibliography{paper}

%

\end{document}